\documentclass{PoS}

\usepackage{pgf,pgfarrows,pgfnodes}

\title{A new approach to the two-dimensional $\sigma$-model with a topological charge}

\ShortTitle{A new approach to the two-dimensional $\sigma$-model}

\author{\speaker{Christian Torrero}$^a$, Oleg Borisenko$^b$, Vladimir Kushnir$^b$, Bartolome All\'es$^c$, Alessandro Papa$^d$\\
        \\   
        \llap{$^a$} Universit\`a di Parma \& I.N.F.N., Parma, Italy\\
        \llap{$^b$} Bogolyubov Institute for Theoretical Physics, Kiev, Ukraine\\
        \llap{$^c$} I.N.F.N., Pisa, Italy\\
        \llap{$^d$} Universit\`a della Calabria \& I.N.F.N., Cosenza, Italy\\
        \\
        E-mail: \email{christian.torrero@fis.unipr.it},\\
                \hspace*{1.1cm} \email{oleg@bitp.kiev.ua},\\
                \hspace*{1.1cm} \email{vkushnir@bigmir.net},\\
                \hspace*{1.1cm} \email{alles@df.unipi.it},\\ 
                \hspace*{1.1cm} \email{papa@fis.unical.it}}

\abstract{Based on character decomposition, a dual transformation is introduced leading to two formulations of the theory which should allow for a removal/softening of the sign problem in the original version. Very preliminar numerical results are commented and remaining problems discussed.}

\FullConference{31st International Symposium on Lattice Field Theory - LATTICE 2013\\
		July 29 - August 3, 2013\\
		Mainz, Germany}

\begin{document}

\section{Introduction}

The two-dimensional ($2D$) non-linear $\sigma$-model~\cite{Polyakov75,Brezin76} has been studied since long given that, among others, it can be related to superconductivity and quantum Hall effect in condensed-matter physics~\cite{Fradkin91} and that it shares common features with non-Abelian gauge theories, such as asymptotic freedom, instantons and spontaneous generation of mass. With the insertion of a $\theta-$term, its action $S_{O(3)}(\beta_{O(3)},\theta)$ in the continuum reads

\begin{equation}
S_{O(3)}(\beta_{O(3)},\theta) = \frac{1}{2}\beta_{O(3)}\int\!\! d^2x\ \![\partial_{\mu}\vec{\sigma}(x)]^2 - i\theta S_q  \ ,
\end{equation}
\vspace*{-0.25cm}

\noindent $\beta_{O(3)}$ being the inverse of the coupling constant, $\theta$ a real parameter, $\vec{\sigma}(x)$ a $3-$component unit vector and $S_q$ the topological charge given by

\begin{equation}
S_q = \frac{1}{8\pi}\int\!\! d^2x \ 
\epsilon^{\mu\nu}\!\epsilon^{kmp}\ \!\partial_{\mu}\sigma_k(x)\ \!\partial_{\nu}\sigma_m(x)\ \!\sigma_p(x)\ .
\end{equation}
\vspace*{-0.25cm}
 
\indent The spectrum of the model displays a massive triplet of scalars~\cite{Hasenfratz90} at $\theta=0$. For $\theta\neq0$, a singlet and triplet exist and, according to the Haldane conjecture~\cite{Haldane77}, they become massless at $\theta=\pi$. Moreover, near $\theta=\pi$ the masses $m_S(\theta)$ and $m_T(\theta)$ of singlet and triplet respectively are both proportional to $(\pi-\theta)^{2/3}$~\cite{Affleck89}. This scenario has been verified --- \emph{either for the triplet only or in the region $\theta\approx\pi$} --- with various techniques that alleviate the sign problem associated with the $S_q$ term~\cite{Bietenholz95,Alles08,deForcrand12,Azcoiti12}.\\
\indent The aim of our study is to allow for numerical simulations with generic real values of an angle $\theta$ so to monitor the behaviour of $m_S(\theta)$ and $m_T(\theta)$ for any $\theta$: this should be accomplished thanks to a dual transformation relating the model in Eq.(1.1) with the $2D$ unconstrained $SU(2)$ principal chiral model described in the following sections.  



\section{The dual formulation - part I}

In order to obtain the above-mentioned duality relation, let us first introduce the more familiar $2D$ $SU(2)$ principal chiral model whose lattice partition function $Z_{SU(2)}(\beta)$ is given by\footnote{As a convention, labels ``$x$" and ``$n$" will denote position in the continuum and on the lattice respectively.}

\vspace*{-0.01cm}
\begin{equation}
Z_{SU(2)}(\beta) = \int \prod_{n} DU (n)\
\exp \left (\! \beta \sum_{n'}\! \sum_{\mu=1}^2  
{\mbox Tr} [U(n')U^{\dagger}(n'+\vec{e}_{\mu})]\!  \right ) \ , 
\end{equation}  
\vspace*{-0.01cm}

\noindent where $\beta$ is the coupling, $U(n)\in SU(2)$ and $n=(n_1,n_2)$ with $n_1,n_2\in\{1,\ldots,L\}$\footnote{Periodic boundary conditions will be assumed from here on.}. $Z_{SU(2)}(\beta)$ can be conveniently rewritten by introducing the link and plaquette variables $V(n,\mu)$ and $V(n)$ defined as

\vspace*{-0.4cm}
\begin{eqnarray}
V(n,\mu) &=& U(n)\ \!U^{\dagger}\!(n+\vec{e}_{\mu})\ ,\\
V(n)\ \ \ \ \ \!&=& V(n,1)\ \!V(n,2)\ \!V^{\dagger}\!(n-\vec{e}_1,1)\ \!V^{\dagger}\!(n-\vec{e}_2,2)\ ,
\end{eqnarray}

\nopagebreak\noindent $V(n)$ being parametrised as $V(n)= \exp [i \lambda_k \omega_k\!(n)]$, with $\left [ \lambda_k,\lambda_m \right ] = 2i\epsilon^{kmp} \lambda_p$ and ${\rm Tr}(\lambda_k\lambda_m) = 2 \delta_{km}$.\\
\newpage
\indent With these definitions, $Z_{SU(2)}(\beta)$ can be rewritten as

\vspace*{-0.1cm}
\begin{equation}
Z_{SU(2)}(\beta) =  \int \prod_{(n,\mu)} dV(n,\mu)\ \!\exp\!{\left [\beta \sum_{(n,\mu)} {\rm Tr} V(n,\mu)\right ]}\prod_{n'} \left ( \sum_r d(r)\chi_r[V(n')] \right )  \ ,
\end{equation}
\vspace*{-0.05cm}

\noindent where the index $r$ labels the representation, $d(r)$ stands for the dimension of the representation $r$ and $\chi_r[V(n)]$ is the character of $V(n)$ in the representation $r$. For future convenience, let us introduce also the unconstrained $SU(2)$ model defined as

\vspace*{-0.05cm}
\begin{equation}
Z(\beta,R) =  \int \prod_{(n,\mu)} dV(n,\mu)\ \!\exp\!{\left [\beta \sum_{(n,\mu)} {\rm Tr} V(n,\mu)\right ]}
\!\prod_{n'} \frac{\sin R \omega(n')}{\sin\omega(n')} \ ,  
\end{equation}

\noindent where $\omega(n)=[\sum_{k=1}^3\omega_k^2(n)]^{\frac{1}{2}}$ and $R$ being a real parameter.\\
\indent The continuum version of the $SU(2)$ principal chiral model stems from the limit $\beta\rightarrow+\infty$, where all link matrices perform small fluctuations around the identity~\cite{Bricmont81}. This allows to replace the $SU(2)$ $\delta-$function with the Dirac $\delta-$function, i.e.,

\begin{equation} 
\sum_rd(r)\chi_r[V(n)] \ \longrightarrow \ 
\prod_{k=1}^{3} \ \int_{-\infty}^{\infty} \ e^{\ \!i\alpha_k(n)\omega_k(n)} \ d\alpha_k(n) \ .
\end{equation}
\vspace*{-0.1cm}

\noindent The continuum limit is then achieved within the following 3-step procedure:

\begin{itemize}

\item let us introduce dimensional vector potentials $\ \!A_k(n)$ as $\omega_k(n)=a A_k(n)$ and expand them in powers of the lattice spacing $a$;

\item substitute the $SU(2)$ invariant measure with a flat measure and extend the integration region over 
potentials $\ \!A_k(n)$ to the non-compact region $\ \!A_k(n)\in[-\infty,\infty]$;

\item in the limit $a\to 0$, replace finite differences with derivatives and sums with integrals.

\end{itemize}

\vspace*{0.15cm}
\indent After some algebra and another change of variables reading

\begin{equation}
\alpha_k(x) = R(x) \ \sigma_k(x) \ , \ \ \ \ \ \ \ \ \ \sum_{k=1}^3\sigma_k^2(x) = 1 \ , 
\end{equation} 
\vspace*{-0.1cm}

\noindent the continuum limit of $Z_{SU(2)}(\beta)$ is finally given by

\vspace*{-0.3cm}
\begin{eqnarray} 
Z_{SU(2)}(\beta) &=&   \int_{0}^{\infty} \ 
\prod_x \ \frac{R^2(x) dR(x)}{\beta \left ( \beta^2 + R^2(x) \right )} \ 
\int \prod_x \left [ \delta\!\left ( 1-\sum_{k=1}^3\sigma_k^2(x) \right ) 
\prod_{k=1}^{3} d\sigma_k(x) \right ]\cdot \nonumber  \\ 
&\cdot&  \exp \left [\! -\int\! d^2x \ {\cal{L}}[R(x),\sigma_k(x)]  \right ]  \ ,  
\end{eqnarray} 
\vspace*{-0.45cm}

\noindent where

\vspace*{0.01cm}
\begin{equation} 
{\cal{L}}[R(x),\sigma_k(x)] \ \equiv \ \frac{1}{4} \ 
\partial_{\mu}[R(x)\sigma_k(x)] \ M_{\mu\nu}^{km}(x)\ \! 
\partial_{\nu}[R(x)\sigma_m(x)] \ , 
\label{L_su2}
\end{equation} 
\vspace*{-0.1cm}

\noindent and

\vspace*{-0.35cm}
\begin{equation} 
M_{\mu\nu}^{km}(x) \ = \ \frac{1}{\beta^2 + R^2(x)} \ 
\left [ \delta_{\mu\nu}\!\left (\! \beta\ \!\delta_{km} + \frac{R^2(x)}{\beta}\sigma_k(x)\ \!\sigma_m(x) 
\right )\! + i\ \!R(x)\ \! \epsilon^{\mu\nu} \epsilon^{kmp} \sigma_p(x)\right ] \ .
\end{equation} 
\vspace*{-0.1cm} 
 
\indent If in Eq.(2.8) $R(x)$ is made independent of $x$ and the integration over this variable is skipped, then not only the continuum limit of $Z(\beta,R)$ in Eq.(2.5) will be obtained, but it can also be proven that the partition function $Z_{O(3)}(\beta_{O(3)},\theta)$ of the initial $2D$ non-linear $\sigma$-model is related to $Z(\beta,R)$ itself via

\vspace*{0.03cm}
\begin{equation} 
Z_{O(3)}(\beta_{O(3)},\theta) = \left [ C(\beta,R) \right ]^{L^2} Z(\beta,R) \ , 
\end{equation}
\vspace*{-0.2cm}

\noindent with

\vspace*{-0.15cm}
\begin{equation} 
C(\beta,R) = \frac{\beta}{R} \left ( R^2+\beta^2 \right ) e^{-2\beta}\ . 
\end{equation} 
\vspace*{-0.15cm}

\noindent The relations between the pairs of parameters $(\beta_{O(3)},\theta)$ and $(\beta,R)$ are

\begin{equation}
\beta_{O(3)} = \frac{\beta}{2} \ \frac{R^2}{R^2 +\beta^2} \ , \ \ \ \ \ \ \ \ \ \ \ 
\theta = 2 \pi R \ \frac{R^2}{R^2 +\beta^2} \ . 
\end{equation} 
\vspace*{-0.12cm} 

\indent Consequently, in order to measure a given observable $O(\sigma)$ in the $2D$ non-linear $\sigma$-model, it is sufficient to ``translate" it into its counterpart $\tilde{O}(V)$ expressed in terms of the degrees of freedom of the $2D$ unconstrained principal chiral model, tune $\beta$ and $R$ so to \emph{keep $\beta$ large}, but in such a way that they correspond to the desired values of $\beta_{O(3)}$ and $\theta$, measure $\tilde{O}(V)$ by means of importance sampling and convert back to $O(\sigma)$.\\ 
\indent As far as the last step is concerned, numerical simulations become now easier, at least in principle. Indeed, Eq.(2.5) reveals that the measure of $Z(\beta,\theta)$ is not complex any more, but it is not necessarily positive yet due to the fluctuating sine functions it contains. Anyway, configurations with negative weight are suppressed in the continuum limit. Bearing this in mind, the strategy we decided to adopt in our preliminary runs is to generate configurations with a standard Metropolis algorithm, automatically rejecting those changes entailing a negative Boltzmann weight. Although this procedure introduces a bias, the above-mentioned suppression makes the systematic errors so generated quite small \cite{Borisenko05,Bricmont81}. 



\section{The dual formulation - part II}

Actually a second formulation of $Z(\beta,R)$ with a real and positive probability distribution exists and has been determined in order to estimate to which extent the above-mentioned bias affects the measurements performed with the algorithm outlined in Section~2. Unfortunately, in this second formulation lattice correlators $G(n_1,n_2)$ - and thus any mass  - are quite onerous to be computed: thus, our strategy will consist in measuring $G(n_1,n_2)$ with the first formulation of $Z(\beta,R)$ only, while other observables\footnote{See the next section for some examples.} will be computed within both approaches for comparison purposes.\\ 
\indent In order to obtain the alternative formulation of $Z(\beta,R)$, let us consider $Z_{SU(2)}(\beta)$ in Eq.(2.4) again and assume that a given representation $r$ has been chosen for all $SU(2)$ matrices so that the partition function $\ \!\tilde{\!Z}(\beta,R)$ --- with $R =2r+1$ --- defined as

\vspace*{-0.1cm}
\begin{equation}
\tilde{\!Z}(\beta,R) = \int\!\! \prod_{(n,\mu)}\!\! dV\!(n,\mu)\ \! \exp{\!\left [\beta \sum_{(n,\mu)} {\rm Tr} V(n,\mu)\right ]}
\!\prod_{n'} \chi_r[V(n')]  \ ,
\end{equation}
\vspace*{-0.1cm}

\noindent can be introduced. It can be shown that

\vspace*{-0.05cm}
\begin{equation}
Z(\beta,R) \ = \ \tilde{\!Z}(\beta,R)\ ,
\end{equation}
\vspace*{-0.3cm}

\noindent when $R$ appearing on the l.h.s. is \emph{integer}. By definition, the character $\chi_r[V(n)]$ reads

\vspace*{0.1cm}
\begin{equation}
\chi_r[V(n)] = \sum_{m_1,m_2,m_3,m_4=-r}^r V(n,1)_{m_1m_2}V(n,2)_{m_2m_3}V^{\dagger}(n-\vec{e}_1,1)_{m_3m_4}V^{\dagger}(n-\vec{e}_2,2)_{m_4m_1}\ ,
\end{equation}
\vspace*{-0.1cm}



\noindent Therefore,\ $\tilde{\!Z}(\beta,R)$ can be rewritten as

\vspace*{0.05cm}
\begin{equation}
\tilde{\!Z}(\beta, R) = \sum_{\{m_1,m_2,p_1,p_2\}=-r}^r\ \prod_{(n',\mu)} Q_{m_1m_2p_1p_2} (n',\mu,\beta)\ ,
\end{equation} 

\noindent with

\vspace*{-0.05cm}
\begin{equation}
Q_{m_1m_2p_1p_2}(n',\mu,\beta) = \int\! dV(n',\mu)\ {\rm e}^{\ \!\!\beta Tr V(n',\mu)}\ \!\!V(n',\mu)_{m_1m_2}
V^{\dagger}(n',\mu)_{p_1p_2}\ .
\end{equation}
\vspace*{-0.1cm} 
 
\noindent Dropping the dependence on $(n,\mu)$ to ease the notations, this last quantity becomes

\vspace*{0.1cm}
\begin{equation}
Q_{m_1m_2p_1p_2}(\beta) = \frac{1}{2 r+1}\sum_J^{2 r} \sum_{k=-J}^J C_J(\beta)
C_{r m_1,\ \!\!Jk}^{\ \!\!r p_2}\ C_{r m_2,\ \!\!Jk}^{\ \!\!r p_1}\ ,
\end{equation}
\vspace*{0.1cm}

\noindent where $C_{r m_1,\ \!\!Jk}^{\ \!\!r p_2}$ are Clebsch-Gordan coefficients and

\vspace*{0.1cm}
\begin{equation}
C_J(\beta) \ = \ \frac{2J+1}{\beta} I_{2J+1}(2\beta) \ ,
\end{equation}
\vspace*{0.01cm}

\noindent $I_{2J+1}(2\beta)$ being modified Bessel functions. Note that the Boltzmann weight in Eq.(3.4) is now \emph{real and positive}. Since

\vspace*{0.1cm}
\begin{equation}
\sum_k C_{r m_1,\ \!\!Jk}^{\ \!\!r p_2}C_{r m_2,\ \!\!Jk}^{\ \!\!r p_1} = C_{r m_1,\ \!\!J (p_2-m_1)}^{\ \!\!r p_2}
C_{r m_2,\ \!\!J (p_1-m_2)}^{\ \!\!r p_1} \delta_{p_2-m_1\!,p_1-m_2}\ ,
\end{equation}
\vspace*{0.05cm}

\noindent just $2$ of the $4$ magnetic numbers associated to each link are eventually free and the count of d.o.f. is restored. Figure 1 shows an example of allowed configuration and how magnetic momenta can be associated either to lattice sites --- as meant in Eq.(3.3) --- or to lattice links --- as understood in Eq.(3.8).\\
\indent In this second formulation, configurations are generated by introducing a discontinuity in an allowed state and letting it propagate randomly until it is reabsorbed. The new configuration is accepted/rejected by a Metropolis test. 

\begin{figure}
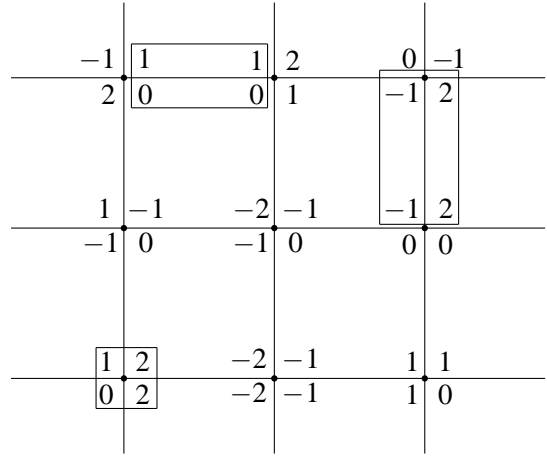

\vspace*{-1.0cm}
\begin{pgfpicture}{-2.8cm}{0cm}{10cm}{10.0cm}
\pgfline{\pgfxy(1.2,4.2)}{\pgfxy(8.3,4.2)}
\pgfline{\pgfxy(1.2,6.2)}{\pgfxy(8.3,6.2)}
\pgfline{\pgfxy(1.2,8.2)}{\pgfxy(8.3,8.2)}
\pgfline{\pgfxy(2.7,3.2)}{\pgfxy(2.7,9.2)}
\pgfline{\pgfxy(4.7,3.2)}{\pgfxy(4.7,9.2)}
\pgfline{\pgfxy(6.7,3.2)}{\pgfxy(6.7,9.2)}
\pgfcircle[fill]{\pgfxy(2.7,4.2)}{1.2pt}
\pgfcircle[fill]{\pgfxy(4.7,4.2)}{1.2pt}
\pgfcircle[fill]{\pgfxy(6.7,4.2)}{1.2pt}
\pgfcircle[fill]{\pgfxy(2.7,6.2)}{1.2pt}
\pgfcircle[fill]{\pgfxy(4.7,6.2)}{1.2pt}
\pgfcircle[fill]{\pgfxy(6.7,6.2)}{1.2pt}
\pgfcircle[fill]{\pgfxy(2.7,8.2)}{1.2pt}
\pgfcircle[fill]{\pgfxy(4.7,8.2)}{1.2pt}
\pgfcircle[fill]{\pgfxy(6.7,8.2)}{1.2pt}
\pgfputat{\pgfxy(2.1,8.3)}{\pgfbox[left,bottom]{$-1 \ \ \ 1$}}
\pgfputat{\pgfxy(2.4,7.85)}{\pgfbox[left,bottom]{$2 \ \ \ \ \!\!0$}}
\pgfputat{\pgfxy(4.37,8.3)}{\pgfbox[left,bottom]{$1 \ \ \ 2$}}
\pgfputat{\pgfxy(4.37,7.85)}{\pgfbox[left,bottom]{$0 \ \ \ 1$}}
\pgfputat{\pgfxy(6.4,8.3)}{\pgfbox[left,bottom]{$0 \ \ \!-\!\!1$}}
\pgfputat{\pgfxy(6.15,7.85)}{\pgfbox[left,bottom]{$-1 \ \ \ \!2$}}
\pgfputat{\pgfxy(2.35,6.3)}{\pgfbox[left,bottom]{$1 \ \ \!-\!1$}}
\pgfputat{\pgfxy(2.15,5.85)}{\pgfbox[left,bottom]{$-1 \ \ \ \ \!\!\!0$}}
\pgfputat{\pgfxy(4.15,6.3)}{\pgfbox[left,bottom]{$-2 \ \ \!\!-\!1$}}
\pgfputat{\pgfxy(4.15,5.85)}{\pgfbox[left,bottom]{$-1 \ \ \ \!0$}}
\pgfputat{\pgfxy(6.15,6.3)}{\pgfbox[left,bottom]{$-1 \ \ \ \!2$}}
\pgfputat{\pgfxy(6.4,5.85)}{\pgfbox[left,bottom]{$0 \ \ \ 0$}}
\pgfputat{\pgfxy(2.37,4.3)}{\pgfbox[left,bottom]{$1 \ \ \ 2$}}
\pgfputat{\pgfxy(2.37,3.85)}{\pgfbox[left,bottom]{$0 \ \ \ 2$}}
\pgfputat{\pgfxy(4.15,4.3)}{\pgfbox[left,bottom]{$\!-2 \ \ \!-\!1$}}
\pgfputat{\pgfxy(4.15,3.85)}{\pgfbox[left,bottom]{$\!-2 \ \ \!-\!1$}}
\pgfputat{\pgfxy(6.44,4.3)}{\pgfbox[left,bottom]{$1 \ \ \ \!1$}}
\pgfputat{\pgfxy(6.44,3.85)}{\pgfbox[left,bottom]{$1 \ \ \ \!0$}}
\pgfline{\pgfxy(2.33,3.8)}{\pgfxy(2.33,4.61)}
\pgfline{\pgfxy(2.33,4.61)}{\pgfxy(3.14,4.61)}
\pgfline{\pgfxy(3.14,4.61)}{\pgfxy(3.14,3.8)}
\pgfline{\pgfxy(3.14,3.8)}{\pgfxy(2.33,3.8)}
\pgfline{\pgfxy(2.8,7.8)}{\pgfxy(2.8,8.65)}
\pgfline{\pgfxy(2.8,8.65)}{\pgfxy(4.61,8.65)}
\pgfline{\pgfxy(4.61,8.65)}{\pgfxy(4.61,7.8)}
\pgfline{\pgfxy(4.61,7.8)}{\pgfxy(2.8,7.8)}
\pgfline{\pgfxy(6.1,6.25)}{\pgfxy(6.1,8.3)}
\pgfline{\pgfxy(6.1,8.3)}{\pgfxy(7.15,8.3)}
\pgfline{\pgfxy(7.15,8.3)}{\pgfxy(7.15,6.25)}
\pgfline{\pgfxy(7.15,6.25)}{\pgfxy(6.1,6.25)}
\end{pgfpicture}
\vspace*{-3.0cm}
\caption{An example of allowed configuration within the second dual formulation of $Z(\beta,R)$ with $R=5$ (i.e., $r=2$). The square and the rectangles show how magnetic momenta can be related either to sites or to links.}
\label{fig1}
\end{figure}


\section{Preliminary results and conclusions}

To test the correctness of the two approaches outlined before, the following two observables 

\begin{equation}
O_1(\beta,R) = \frac{\partial\ln[Z(\beta,R)]}{\partial\beta}\ , \ \ \ \ \ \ \ \ \ \ \ O_2(\beta, R, J) = \langle\sum_{k=-J}^J C_{r m_1,\ \!\!Jk}^{\ \!\!r p_2}C_{r m_2,\ \!\!Jk}^{\ \!\!r p_1}\rangle\ ,
\end{equation}
\vspace*{-0.1cm}

\noindent have been numerically computed in the strong-coupling regime and compared to available analytical results. Concerning $O_2(\beta,J)$, it is important to stress that it has no particular physical meaning and that it can be computed just in the formulation described in Section 3 because of its definiton.\\
\indent Table 1 shows how numerical estimates compare to analytical values: not only the agreement is good in general but, at least in this regime and for these quantities, the bias affecting the first formulation does not apparently impact too much on the data.\\ 
\indent After these encouraging results, the next test being performed has been an attempt to recover a periodic signal for suitable observables for values of the parameters corresponding to a fixed $\beta_{O(3)}$ and varying $\theta$ in the initial non-linear $\sigma$-model. 

\begin{table}
\begin{center}
\begin{tabular}{|l|c|c|c|}
\hline
$Observable$ & $Strong-coupling\ prediction$ & $1^{st} formulation$ & $2^{nd} formulation$ \\
\hline
$O_1(0.1,7)$ & $0.09983$ & $0.100(18)$ & $0.0999(6)$ \\
\hline
$O_1(0.3,11)$ & $0.29560$  & $0.296(17)$ & $0.2956(17)$ \\
\hline
$O_1(0.5,15)$ & $0.48039$ & $-$ & $0.4804(26)$ \\  
\hline
$O_1(0.7,15)$ & $0.64918$ & $0.649(16)$ & $-$  \\  
\hline
$O_2(0.3,11,1)$ & $0.00498$ & $-$ & $0.0049(61)$ \\  
\hline
$O_2(0.5,15,1)$ & $0.01308$ & $-$ & $0.0131(60)$ \\  
\hline
\end{tabular}
\caption{Comparison of analytical values of $O_1(\beta,R)$ and $O_2(\beta, R, J)$ with numerical estimates evaluated with both formulations of $Z(\beta, R)$ with $L=40$: $1^{st}$ and $2^{nd}$ formulation stand for $Z(\beta,R)$ as written in Eq.(2.5) and Eq.(3.4) respectively. $O_2(\beta, R, J)$ can actually be measured just within the approach described in Section 3.}
\label{tab1}
\end{center}
\end{table}


\end{document}